\documentclass[12pt]{article}
\usepackage{graphicx,epsfig}
\usepackage{amssymb}

\begin{document}
\begin{center}
{\Large Formation of Light Isotopes by Protons and Deuterons of \\
3.65 GeV/nucleon on Separated Tin Isotopes} \\

\vspace{8mm} 
A. R. Balabekyan$^{1,2}$, 
A. S. Danagulyan$^{1}$,
J. R. Drnoyan$^1$, 
G. H. Hovhannisyan$^1$,\\
J. Adam$^{2,3}$, V. G. Kalinnikov$^2$, M. I. Krivopustov$^2$,
V. S. Pronskikh$^2$, V. I. Stegailov$^2$,\\
A. A. Solnyshkin$^2$, P. Chaloun$^{2,3}$, V. M. Tsoupko-Sitnikov$^2$\\
S. G. Mashnik$^4$, K. K. Gudima$^5$ \\

\vspace{2mm} 
$^1${\it Yerevan State University, Armenia} \\
$^2${\it JINR, Dubna, Russia} \\ 
$^3${\it INF AS R\v{e}z, Czech Republic} \\
$^4${\it Los Alamos National Laboratory, Los Alamos, NM 87545, USA} \\
$^5${\it Institute of Applied Physics, Academy of Science of
Moldova, Chi\c{s}in\u{a}u}
\end{center}

\begin{center}
{\bf Abstract}
\end{center}

We measure cross sections for residual nuclide formation in the mass 
range $7\leq A \leq 96$ caused by bombardment with protons and 
deuterons of 
3.65 GeV/nucleon energy of enriched tin isotopes ($^{112}{\rm Sn},
^{118}{\rm Sn}, ^{120}{\rm Sn}, ^{124}{\rm Sn}$). The experimental
data are compared with calculations by the
codes FLUKA, LAHET, CEM03, and LAQGSM03.
Scaling behavior is observed for the whole mass region of 
residual nuclei, showing a possible multifragmentation
mechanism for the formation of light products ($7\leq A \leq 30$). Our
analysis of the isoscaling dependence also shows 
a possible contribution of multifragmentation to the production 
of heavier nuclides, in the mass region $40\leq A \leq 80$.

\section{Introduction}

The nuclear reaction mechanism of fragmentation has been investigated
for more than 60 years. A turning point in this study was marked by 
Jacobsson {\it et al.} in 1982, who measured multiple fragment 
production in nuclear emulsions containing Ag and Br, irradiated with
$^{12}{\rm C}$ at 55$A$ and 110$A$ MeV/nucleon \cite{jac}. 
These data stimulated development of new models to explain the
formation of multiple fragments by a ``liquid-gas" phase transition
in hot nuclear matter (see, {\it e.g.}, \cite{bot1,ban}).

The isospin dependence in the equation of state of nuclear matter is
very important, being at the same time poorly known property of 
neutron-rich nuclear matter \cite{bao}. In recent years, much 
attention has been paid to the isospin dependence both in 
nucleus-nucleus experiments with an excess of neutrons in the 
bombarding and/or target nuclei
and in experiments with different types of light projectiles on 
targets with different neutron/proton ratios 
\cite{fri}--\cite{mat}.
Such investigations may help obtain information
about the equation of state of the asymmetric nuclear matter.

Many experiments have been devoted to the study of
nuclear multifragmentation, where several fragments in the mass
region $3 \leq Z \leq 20$ are formed from hot nuclear matter 
\cite{rod,kar}. 
Observation of isoscaling, that is, the dependence of fragment
formation probabilities on the third component of their isotopic spins, 
has increased the possibility of obtaining  information on the 
formation mechanisms of these fragments \cite{tsang1}--\cite{tsang2}. 
Early work in this field was done by Bogatin {\it et al.} \
\cite{bogat1,bogat2} and has continued \cite{bot2}.

Recently experimentalists and  theorists have focussed on the
investigation of formation mechanisms of
heavy fragments ($Z\geq 20$) by different
projectiles ($\gamma$ - rays, $\pi^-$- meson etc.) 
\cite{fri}--\cite{mat}. 
Experiments with a direct registration of heavy fragments 
do not provide a comprehensive understanding 
of fragments in this mass region.
The induced-radioactivity method adds more
possibilities and the
investigation of mechanisms of heavy-fragment production 
becomes more realistic \cite{bal3,bal2}.

The aim of the present work is to investigate the formation of
product nuclei on separated tin isotopes by
proton and deuteron beams of 3.65 GeV/nucleon
in three mass regions of product nuclides: $7 \leq A \leq 30$, $40 \leq
A \leq 80$, and $81 \leq A \leq 96$.

\section{Experimental results and discussion}

Targets of enriched tin isotopes $^{112}{\rm Sn}, ^{118}{\rm Sn}, 
^{120}{\rm Sn}$, and
$^{124}{\rm Sn}$ are irradiated at the Nuclotron and Synchrophasotron
of the LHE JINR in Dubna by proton and deuteron beams with energy 
of 3.65 GeV/nucleon. The description of the experiment is given 
in \cite{bal1}.
The measurement of products in the mass range  $7\leq A \leq 80$
is performed by studying the induced activity. The radioactive
nuclei obtained are identified by the characteristic $\gamma$-rays and
their half-lives.  For beam monitoring, we employ the reactions
$^{27}$Al$(d,3p2n)^{24}$Na  and  $^{27}$Al$(p,3pn)^{24}$Na, whose cross 
sections are taken as of $14.2\pm0.2$ mb \cite{8} and  
$10.6\pm0.8$ mb \cite{9}, respectively.
Cross sections for about 70 products from each target
for the proton and deuteron beams are obtained.
The measured cross sections of all products are shown in 
Tables 1 and 2: index ``I" denotes independent yields while ``C" indicates
cumulative ones.

\begin{center}
\textbf{Table 1}. Measured product cross sections 
for p + $^{112,118,120,124}$Sn

\vspace*{5mm}
\begin{tabular}{|c|c|c|c|c|c|}  \hline
Product & Type &
\multicolumn{4}{|c|}{Cross section (mb)} \\ \cline{3-6}
&&$^{112}{\rm Sn}$&$^{118}{\rm Sn}$&$^{120}{\rm Sn}$&$^{124}{\rm Sn}$ \\ \hline

$^7$Be    &I& 13.9$\pm$1.5  & 9.4$\pm$0.3   & 8.2$\pm$ 1.4  & 7.5$\pm$ 0.8 \\
\hline 
$^{22}$Na &C&  2.3$\pm$0.3  & 2.4$\pm$0.4   & 2.1$\pm$ 0.4  & 1.7$\pm$0.2 \\ 
\hline
$^{24}$Na &C& 3.25$\pm$0.3  & 3.23$\pm$0.2  & 3.69$\pm$0.3  & 3.97$\pm$0.3 \\ 
\hline
$^{28}$Mg &C& 0.39$\pm$0.06 & 0.53$\pm$0.05 &0.75$\pm$0.08  & 0.89$\pm$0.07 \\ 
\hline 
$^{38}$Cl &I&               & 1.67$\pm$ 0.2 &1.5$\pm$ 0.2   &               \\
\hline 
$^{39}$Cl &C&               & 0.57$\pm$0.02 & 0.34$\pm$0.07 &               \\
\hline 
$^{42}$K  &C& 1.76$\pm$0.11 & 1.85$\pm$0.25 & 1.95$\pm$0.2  & 2.1$\pm$0.2   \\
\hline 
$^{43}$K  &C&  0.74$\pm$0.06& 0.85$\pm$ 0.08& 1.04$\pm$ 0.1 & 1.32$\pm$ 0.1 \\
\hline 
$^{43}$Sc &C& 0.72$\pm$0.18 &0.6$\pm$0.2    & 0.45$\pm$0.2  & 0.2$\pm$0.03 \\
\hline 
$^{44g}$Sc&I& 0.97$\pm$ 0.09& 0.54$\pm$0.15 & 0.58$\pm$0.04 & 0.36$\pm$0.09 \\
\hline 
$^{44m}$Sc&I& 2.28$\pm$0.1 & 1.45$\pm$0.06  & 1.4$\pm$0.07  & 1.7$\pm$0.1   \\
\hline
$^{46}$Sc &I& 2.2$\pm$ 0.2 & 2.35$\pm$0.2   & 2.6$\pm$0.4   & 2.4$\pm$0.2  \\
\hline 
$^{48}$Sc &I& 0.3$\pm$ 0.05& 0.38$\pm$0.07  & 0.42$\pm$0.05 & 0.7$\pm$0.09 \\
\hline 
$^{48}$Cr &C& 0.19$\pm$0.07& 0.1$\pm$ 0.01  & 0.13$\pm$ 0.04&             \\
\hline 
$^{51}$Cr &C&              & 3.7$\pm$0.6    & 3.3$\pm$0.6   &             \\
\hline  
$^{48}$V  &I& 2.7$\pm$ 0.15& 1.68$\pm$ 0.1  & 1.76$\pm$ 0.1 & 1.06$\pm$ 0.1 \\
\hline  
$^{52}$Mn &C& 1.9$\pm$0.04 & 1.12$\pm$0.03  & 1.03$\pm$0.04 & 0.7$\pm$0.08 \\
\hline 
$^{54}$Mn &I& 6.1$\pm$0.3  & 5.1$\pm$0.25   & 4.8$\pm$0.3   & 4.2$\pm$0.3 \\ 
\hline 
$^{56}$Mn &C& 0.8$\pm$ 0.03& 1.08$\pm$0.07  & 1.33$\pm$0.1  & 1.86$\pm$0.08 \\
\hline 
$^{59}$Fe &C& 0.37$\pm$0.05& 0.83$\pm$ 0.09 & 0.85$\pm$ 0.07& 1.17$\pm$ 0.1 \\
\hline 

\end{tabular}
\end{center}

\begin{center}
\textbf{Table 1} (continued)

\vspace*{5mm}
\begin{tabular}{|c|c|c|c|c|c|}  \hline
Product & Type &
\multicolumn{4}{|c|}{Cross section (mb)} \\ \cline{3-6}
          & &$^{112}$Sn     &$^{118}$Sn     &$^{120}$Sn     &$^{124}$Sn   \\ 
\hline
$^{56}$Co &C& 1.8$\pm$0.1  & 1.7$\pm$ 0.2   & 1.75$\pm$ 0.2 & 1.54$\pm$0.2 \\
\hline 
$^{58}$Co &C& 5.7$\pm$ 0.4 &4.8$\pm$ 0.5    & 4.8$\pm$ 0.4  & 3.6$\pm$0.4 \\
\hline 
$^{67}$Cu &C& 0.1$\pm$0.02 & 0.29$\pm$0.05  & 0.44$\pm$0.04 & 0.52$\pm$0.05 \\
\hline 
$^{65}$Zn &C& 9.1$\pm$0.3  & 6.9$\pm$0.3    & 6.9$\pm$0.3   & 5.1$\pm$0.3 \\ 
\hline 
$^{66}$Ga &C& 4.8$\pm$0.4  & 3.1$\pm$0.2    & 2.9$\pm$0.3   & 2.5$\pm$0.3  \\ 
\hline 
$^{67}$Ga &C& 8.9$\pm$0.07 &6.7$\pm$ 0.4    & 6.4$\pm$ 0.4  & 5.3$\pm$ 0.5 \\
\hline 
$^{69}$Ge &C& 7.5$\pm$ 0.7 &5.3$\pm$ 0.5    & 5.0$\pm$ 0.3  & 3.9$\pm$ 0.5 \\
\hline 
$^{77}$Ge &C&              & 0.2$\pm$ 0.05  & 0.16$\pm$ 0.05&               \\
\hline 
$^{70}$As &C& 3.0$\pm$ 0.5 & 1.9$\pm$ 0.5   & 1.42$\pm$ 0.2 & 2.1$\pm$ 0.6 \\
\hline 
$^{71}$As &C& 8.01$\pm$0.8 & 5.54$\pm$ 0.06 & 5.3$\pm$ 0.4  & 4.1$\pm$ 0.5 \\
\hline 
$^{72}$As &C& 2.8$\pm$ 0.4 & 2.1$\pm$ 0.3   & 1.9$\pm$ 0.4  & 2.1$\pm$ 0.6 \\
\hline 
$^{74}$As &I& 1.92$\pm$0.15& 2.6$\pm$ 0.3   & 3.07$\pm$ 0.4 & 3.5$\pm$ 0.25 \\
\hline 
$^{76}$As &I& 4.5$\pm$ 0.4 & 5.1$\pm$ 0.4   & 5.3$\pm$ 0.5  & 6.3$\pm$ 0.4 \\
\hline 
$^{73}$Se &C& 5.7$\pm$ 0.15& 3.8$\pm$ 0.15  & 3.5$\pm$ 0.15 & 2.4$\pm$ 0.2 \\
\hline
$^{75}$Se &C& 13.2$\pm$0.5 & 10.3$\pm$ 0.7  & 10.1$\pm$ 1.0 & 8.8$\pm$ 0.7 \\
\hline
$^{76}$Br &C& 10.5$\pm$1   & 7.3$\pm$0.5    & 6.6$\pm$0.7   & 5.2$\pm$0.4  \\
\hline 
$^{77}$Br &I& 10.4$\pm$0.4 & 8.4$\pm$0.2    & 8.4$\pm$0.2   & 7.8$\pm$0.2 \\
\hline
$^{82}$Br &I&              & 0.26$\pm$0.04  & 0.25$\pm$0.02 & 0.49$\pm$0.03 \\
\hline 
$^{76}$Kr &C& 1.6$\pm$ 0.2 & 1.03$\pm$ 0.05 & 0.9$\pm$ 0.07 & 0.55$\pm$0.05 \\
\hline 
$^{77}$Kr &C&              & 3.2$\pm$ 0.3   &2.7$\pm$ 0.3   & 1.76$\pm$0.16\\
\hline
$^{81}$Rb &C& 16.4$\pm$0.6 & 11.7$\pm$0.2   & 11.5$\pm$0.4  & 8.8$\pm$0.6\\ 
\hline 
$^{82m}$Rb&I& 5.2$\pm$0.05 &5.3$\pm$0.4   & 5.9$\pm$0.4  &    5.7$\pm$0.3 \\
\hline 
$^{83}$Rb &C& 18.6$\pm$ 0.7& 15.3$\pm$ 0.5& 16.0$\pm$ 0.6  & 13.8$\pm$ 0.3 \\
\hline
$^{84g}$Rb&I& 1.4$\pm$ 0.3 & 2.3$\pm$ 0.2 & 2.8$\pm$ 0.2   & 4.4$\pm$ 0.6 \\
\hline 
$^{86}$Rb &I& 0.34$\pm$0.09&              & 1.09$\pm$0.14  & 1.85$\pm$0.24\\
\hline 
$^{83}$Sr &C& 14.6$\pm$0.1 & 10.3$\pm$0.3 & 9.8$\pm$0.2   & 7.4$\pm$0.5\\ 
\hline 
$^{85}$Sr &C& 21$\pm$1.6   & 17.5$\pm$1.7 & 17.3$\pm$1.5  & 15.2$\pm$0.9\\ 
\hline
$^{84m}$Y &I& 5.1$\pm$0.5  & 3.4$\pm$0.3  & 2.6$\pm$0.4   & 2.1$\pm$0.4 \\
\hline 
$^{86m}$Y &I& 4.9$\pm$ 0.4 & 6.4$\pm$ 0.1 & 6.7$\pm$ 0.4  & 5.5$\pm$ 0.4 \\ 
\hline 
$^{87m}$Y &C& 18.6$\pm$ 0.7& 15.5$\pm$ 0.9& 15.8$\pm$ 0.3 & 12.8$\pm$ 0.4 \\ 
\hline 
$^{87g}$Y &I& 4.3$\pm$ 0.4 &3.5$\pm$ 0.3  & 4.0$\pm$ 0.3  & 2.8$\pm$ 0.2 \\ 
\hline 
$^{86}$Zr &C& 8.8$\pm$ 0.5 & 4.8$\pm$ 0.15& 3.5$\pm$ 0.1  & 2.3$\pm$ 0.25 \\
\hline 
$^{88}$Zr &C& 20.2$\pm$2.0 & 13.8$\pm$0.8 &  14.4$\pm$1.0 &  10.2$\pm$0.9 \\
\hline 
$^{89}$Zr &C& 20.2$\pm$0.5 & 16.35$\pm$0.3&  16.4$\pm$0.3 &  13.1$\pm$0.7 \\
\hline
$^{90}$Nb  &C& 18.2$\pm$ 1.0& 12.4$\pm$ 0.4& 11.6$\pm$ 1.2 & 8.6$\pm$ 0.4 \\
\hline
$^{95g}$Nb&C&             &0.8$\pm$ 0.03 & 1.75$\pm$ 0.07& 2.40$\pm$ 0.25 \\
\hline 
$^{95m}$Nb&I&             &0.17$\pm$0.08 & 0.35$\pm$ 0.06&                \\
\hline
$^{96}$Nb &I&0.33$\pm$0.08&0.42$\pm$ 0.07& 0.65$\pm$0.06  & 1.14$\pm$ 0.16 \\
\hline 
$^{90}$Mo &C& 5.9$\pm$0.2 & 2.6$\pm$0.3  & 2.1$\pm$ 0.3  & 1.1$\pm$ 0.1 \\
\hline 
$^{93m}$Mo&I& 3.5$\pm$ 0.2&4.1$\pm$ 0.4  & 4.4$\pm$ 0.3  & 3.8$\pm$ 0.2 \\
\hline 
$^{99}$Mo &C& 0.19$\pm$0.02&0.26$\pm$ 0.02& 0.62$\pm$0.13& 1.65$\pm$0.25 \\
\hline 
$^{93}$Tc &C& 12.35$\pm$0.8 &6.95$\pm$0.3 & 5.7$\pm$ 0.5 & 3.4$\pm$ 0.3 \\
\hline
\end{tabular}
\end{center}

\newpage
\begin{center}
\textbf{Table 1} (continued)

\vspace*{5mm}
\begin{tabular}{|c|c|c|c|c|c|}  \hline
Product & Type &
\multicolumn{4}{|c|}{Cross section (mb)} \\ \cline{3-6}
            & & $^{112}$Sn   &   $^{118}$Sn & $^{120}$Sn   & $^{124}$Sn \\ 
\hline
 $^{94}$Tc  &I& 9.8$\pm$ 0.2 & 6.5$\pm$ 0.1 & 6.7$\pm$ 0.2 & 4.4$\pm$ 0.3 \\
\hline 
$^{95g}$Tc  &I& 12.4$\pm$ 0.6&9.8$\pm$ 0.4  & 8.3$\pm$ 0.3 & 7.5$\pm$ 0.3 \\
\hline 
$^{95m}$Tc  &I& 1.0$\pm$0.1  &1.1$\pm$0.1   & 0.8$\pm$0.08 & 0.56$\pm$0.10 \\ 
\hline
$^{96}$Tc   &I& 4.4$\pm$ 0.1 & 6.7$\pm$ 0.2 & 7.4$\pm$ 0.25& 6.4$\pm$0.1 \\ 
\hline
\end{tabular}
\end{center}

\vspace*{5mm}
\begin{center}
\textbf{Table 2}. Measured product cross sections 
for d + $^{112,118,120,124}$Sn

\vspace*{5mm}
\begin{tabular}{|c|c|c|c|c|c|}  \hline
Product & Type &
\multicolumn{4}{|c|}{Cross section (mb)} \\ \cline{3-6}
&&$^{112}$Sn&$^{118}$Sn&$^{120}$Sn&$^{124}$Sn \\ 
\hline 
$^7$Be     &I& 31.1$\pm$ 2.7&              & 25.5$\pm$ 2.5&23.1$\pm$4.0 \\ 
\hline
$^{22}$Na  &C& 23.3$\pm$ 0.4& 8.1$\pm$ 1.5 & 4.1$\pm$ 0.1 & 3.5$\pm$0.9 \\ 
\hline 
$^{24}$Na  &C& 6.2$\pm$ 0.4 & 10.0$\pm$ 1.3& 9.9$\pm$0.8  & 12.2$\pm$ 1.1 \\
\hline 
$^{28}$Mg  &C& 1.0$\pm$ 0.1 & 1.8$\pm$0.1  & 1.6$\pm$ 0.2 & 2.9$\pm$ 0.8 \\ 
\hline 
$^{38}$S   &C&              &              & 0.37$\pm$0.04& 0.56$\pm$0.06 \\
\hline 
$^{38}$Cl  &I&              &              & 3.0$\pm$ 0.2 & 3.5$\pm$ 0.5 \\ 
\hline 
$^{39}$Cl  &C&              &              & 1.1$\pm$ 0.1 & 1.8$\pm$ 0.6\\ 
\hline 
$^{42}$K   &C& 2.4$\pm$ 0.2 & 4.5$\pm$ 1.2 & 3.9$\pm$ 0.5 & 5.2$\pm$ 0.5 \\ 
\hline
$^{43}$K   &C& 1.2$\pm$ 0.1 & 2.4$\pm$ 0.4 & 2.1$\pm$0.3  & 3.1$\pm$ 0.3 \\ 
\hline 
$^{43}$Sc  &C& 1.2$\pm$ 0.1 & 1.4$\pm$0.02 & 1.4$\pm$ 0.1 & 2.1$\pm$ 0.1 \\ 
\hline 
$^{44m}$Sc &I& 2.9$\pm$0.1  & 3.2$\pm$ 0.4 & 2.7$\pm$ 0.7 & 2.0$\pm$ 0.3 \\ 
\hline
$^{44g}$Sc &I& 1.7$\pm$ 0.4 & 2.0$\pm$ 0.3 & 1.5$\pm$ 0.2 & 1.5$\pm$0.2 \\ 
\hline 
$^{46}$Sc  &I& 3.3$\pm$ 0.8 & 6.1$\pm$ 0.8 & 6.1$\pm$0.3  & 6.6$\pm$ 0.3 \\ 
\hline 
$^{47}$Sc  &C& 3.5$\pm$ 0.2 &              &              &               \\
\hline 
$^{48}$Sc &I& 0.5$\pm$ 0.1 & 1.0$\pm$ 0.09 & 1.1$\pm$ 0.2 &1.6$\pm$ 0.3 \\ 
\hline 
$^{48}$V  &C& 3.5$\pm$ 0.3 & 3.5$\pm$ 0.4  & 3.2$\pm$ 0.1 & 2.9$\pm$ 0.5 \\ 
\hline 
$^{51}$Cr &C& 14.1$\pm$ 1.4&6.1$\pm$ 0.4   & 7.4$\pm$0.8  & 5.7$\pm$ 0.5 \\ 
\hline 
$^{52g}$Mn&C& 2.3$\pm$ 0.4 & 2.1$\pm$ 0.2 & 2.0$\pm$ 0.4 & 1.5$\pm$ 0.3 \\
\hline 
$^{56}$Mn &C& 2.4$\pm$ 0.6 & 3.1$\pm$ 0.4 & 2.9$\pm$ 0.1 & 4.3$\pm$ 0.3 \\ 
\hline 
$^{59}$Fe &C& 0.68$\pm$0.03& 1.5$\pm$ 0.2 & 1.7$\pm$ 0.1 & 2.7$\pm$0.2 \\ 
\hline 
$^{55}$Co &C& 0.35$\pm$0.05&             &               &             \\ 
\hline 
$^{56}$Co &C& 1.9$\pm$ 0.1 & 1.9$\pm$ 0.3& 1.4$\pm$ 0.1 & 1.1$\pm$ 0.1 \\ 
\hline 
$^{57}$Co &C& 6.9$\pm$ 0.2 & 11.8$\pm$0.3& 5.9$\pm$ 0.2 & 4.8$\pm$ 0.1 \\
\hline 
$^{58}$Co &I& 8.3$\pm$ 0.3 & 9.9$\pm$ 0.2& 7.6$\pm$ 1.0 & 7.8$\pm$ 0.5 \\ 
\hline  
$^{60}$Cu &C& 1.9$\pm$ 0.4 &1.04$\pm$0.19& 1.7$\pm$ 0.1 & 0.5$\pm$ 0.09 \\ 
\hline 
$^{67}$Cu &C&              &             &0.34$\pm$ 0.04& 0.34$\pm$ 0.01 \\ 
\hline 
$^{62}$Zn &C& 1.25$\pm$0.05&             &              & 0.3$\pm$0.03 \\ 
\hline 
$^{65}$Zn &C& 16.5$\pm$1.0 &             & 10.7$\pm$ 0.2& 10.1$\pm$ 0.4 \\ 
\hline 
$^{69m}$Zn &I&0.39$\pm$ 0.05&0.84$\pm$0.05& 1.1$\pm$ 0.1 & 1.4$\pm$ 0.1 \\
\hline 
$^{66}$Ga &C& 5$\pm$ 0.5   &             &3.3$\pm$0.3   & 3.4$\pm$ 0.3 \\
\hline 
$^{67}$Ga &C& 8.6$\pm$ 0.3 & 9.7$\pm$ 1.3& 9.3$\pm$ 0.7 & 8.7$\pm$ 0.3 \\ 
\hline  
$^{73}$Ga &C& 0.48$\pm$0.12&             &              &               \\ 

\hline
\end{tabular}
\end{center}

\newpage
\begin{center}
\textbf{Table 2} (continued)

\vspace*{5mm}
\begin{tabular}{|c|c|c|c|c|c|}  \hline
Product & Type &
\multicolumn{4}{|c|}{Cross section (mb)} \\ \cline{3-6}
          & & $^{112}$Sn   & $^{118}$Sn   & $^{120}$Sn   & $^{124}$Sn \\ 
\hline 
$^{69}$Ge &C& 6.2$\pm$ 0.6 & 8.8$\pm$ 1.6 &7.2$\pm$ 0.5  & 6.7$\pm$ 0.9 \\ 
\hline 
$^{77}$Ge &C&              &               & 1.3$\pm$0.1 & 2.5$\pm$ 0.3 \\
\hline 
$^{71}$As &C& 7.4$\pm$ 0.2 & 7.73$\pm$0.8  & 6.7$\pm$ 0.2& 5.4$\pm$ 0.4\\ 
\hline
$^{74}$As &I& 2.1$\pm$ 0.1 & 2.65$\pm$ 0.5 & 4.3$\pm$ 0.1& 6.9$\pm$ 0.8 \\ 
\hline  
$^{78}$As &C& 1.35$\pm$0.05&               &             & 0.9$\pm$ 0.2\\ 
\hline 
$^{73g}$Se &C& 6.6$\pm$ 0.1 & 6.4$\pm$ 0.4  & 5.2$\pm$ 0.6& 4.3$\pm$ 0.3 \\
\hline 
$^{75}$Se &I& 7.0$\pm$ 0.77& 9.4$\pm$ 1.49 &10.8$\pm$0.93&10.3$\pm$ 1.06 \\ 
\hline 
$^{75}$Br &C& 6.8$\pm$ 0.6 &5.4$\pm$ 0.5   & 4.4$\pm$ 0.3& 3.5$\pm$ 0.3 \\ 
\hline  
$^{77}$Br &I& 4.5$\pm$ 0.55& 6.9$\pm$ 1.18 & 6.0$\pm$ 0.47& 6.4$\pm$ 0.74 \\
\hline  
$^{77}$Kr &C& 5.2$\pm$ 0.6 & 3.5$\pm$ 0.5  & 3.5$\pm$ 0.2 &2.4$\pm$ 0.2 \\ 
\hline   
$^{79}$Kr &I& 1.2$\pm$ 0.13&               &8.9$\pm$1.22  & 8.0$\pm$ 1.01\\ 
\hline 
$^{79}$Rb &C& 10.1$\pm$1.0 &               &2.8$\pm$0.3   & 1.9$\pm$0.2 \\
\hline 
$^{81}$Rb &C& 14.1$\pm$ 0.7& 15.7$\pm$ 0.6 & 12.8$\pm$ 0.4&10.9$\pm$ 0.7 \\ 
\hline 
$^{83}$Rb &C&20.3$\pm$ 0.8 &24.15$\pm$ 0.95& 21.7$\pm$ 1.0& 21.6$\pm$ 0.6 \\
\hline 
$^{84}$Rb &I& 1.2$\pm$ 0.1 & 3.2$\pm$ 0.3  & 4.1$\pm$ 0.5&6.4$\pm$ 0.3 \\ 
\hline 
$^{82}$Sr &C& 10.1$\pm$1.5 & 8.8$\pm$ 1.5  & 6.0$\pm$0.6 & 4.6$\pm$0.5 \\ 
\hline  
$^{83}$Sr &C& 13.1$\pm$ 0.3&13.7$\pm$ 1.2  & 10.3$\pm$1.3& 9.0$\pm$ 0.6 \\ 
\hline  
$^{85}$Sr &I&              &               &13.3$\pm$1.55& 15.9$\pm$ 1.4 \\ 
\hline 
$^{85m}$Y &C&              &               &7.7$\pm$ 0.9 & 3.5$\pm$ 0.3 \\ 
\hline 
$^{85g}$Y &C&              &               &             &2.9$\pm$ 0.7 \\ 
\hline  
$^{86m}$Y &I& 5.0$\pm$ 0.3 & 8.3$\pm$ 0.2  &7.6$\pm$ 0.1 & 6.6$\pm$ 0.2 \\ 
\hline 
$^{86g}$Y &I& 6.5$\pm$ 0.5 &               & 8.9$\pm$ 0.1& 8.8$\pm$ 0.8 \\ 
\hline 
$^{87g}$Y &C& 20.5$\pm$ 2.1&               &20.0$\pm$ 2.0& 19.3$\pm$ 2.0 \\ 
\hline 
$^{88}$Y  &I&              &10.2$\pm$1.7   & 5.6$\pm$ 0.5& 7.8$\pm$ 0.8 \\ 
\hline 
$^{86}$Zr &C& 7.4$\pm$0.2  & 5.4$\pm$ 0.1  & 3.7$\pm$ 0.3& 2.8$\pm$ 0.1 \\ 
\hline 
$^{88}$Zr &C& 18.4$\pm$ 2.2&               &16.8$\pm$ 0.2& 14.6$\pm$ 0.2\\ 
\hline  
$^{89}$Zr &C& 18.4$\pm$ 0.4&               & 17.8$\pm$ 0.5& 16.9$\pm$ 0.7 \\
\hline
$^{93m}$Tc&I& 10.1$\pm$ 0.6&               & 6.6$\pm$ 0.5 & 2.0$\pm$ 0.2\\
\hline
$^{94m}$Tc&I& 9.5$\pm$ 1.0 &               &              & 5.6$\pm$ 0.2\\
\hline
$^{94g}$Tc&I& 4.3$\pm$ 0.3 &1.9$\pm$0.46   &              & 2.4$\pm$ 0.8\\
\hline
$^{95m}$Tc&I&              &15.1$\pm$0.7   &              &             \\
\hline
$^{95g}$Tc&I& 1.23$\pm$0.05&1.23$\pm$0.07  & 1.2$\pm$0.2  & 1.0$\pm$ 0.1\\
\hline
$^{96g}$Tc&I& 4.4$\pm$ 0.3 &9.1$\pm$0.6    &              & 8.4$\pm$ 1.3\\
\hline

\end{tabular}
\end{center}

To reveal the production mechanisms 
of light nuclei, the experimental results are analyzed from the viewpoint of:

1) exponential dependence of cross sections on the mass and charge numbers;

2) including isospin dependence.

Investigations by many authors have showed that  
the yields of fragments from various nuclear reactions
can be represented as
$\sigma(A_f)\sim
A_{f}^{-\tau}$ and $\sigma(Z_f)\sim Z_{f}^{-\tau}$, 
where  $\tau$ has values of about $1.5$--$2$ depending on 
the reactions, where $A_f$ and $Z_f$ are the mass and charge
numbers of the fragments. 
Note that calculations by the Statistical Multifragmentation Model
(SMM) \cite{ban} for the mass region of
fragments discussed here provide an exponential dependence 
with $\tau=2.2$.

The isospin dependence of the available experimental yields points to
an isoscaling behavior.
In the case of multifragmentation, the ratio
of the yields of fragments produced from different targets has an
exponential dependence on the number of protons and neutrons of the
product isotopes described by the formula \cite{tsang1}:
\begin{equation}
 R_{21}(t_3)=Y_{2}(N,Z)/Y_{1}(N,Z)= C \exp(\alpha N+\beta Z),
\end{equation}
where $Y(N,Z)$ is the yield of fragment with $Z$ protons and  $N$ neutrons,
and  $t_3=(N-Z)/2$ is the third projection of the fragment isospin.
Indices 1 and 2 correspond to different targets with
different isotopic compositions, with 2 corresponding to the more
neutron-rich target and where $C$ is a normalization parameter.
In Ref.\ \cite{xu}, the parameters $\alpha$ and $\beta$ were
expressed using the difference of chemical potentials
of the two systems as following: $\alpha = \Delta \mu_n/T$,
$\beta=\Delta \mu_p/T$, where $T$ is the temperature of the excited nucleus.

Since in our measurements we use targets of different isotopes
of the same element, we analyze our data with the following formula:
\begin{equation}
 R_{21}(t_3)=Y_{2}(N,Z)/Y_{1}(N,Z)= \exp(C+B t_3),
\end{equation}
where
$C$ and $B$ are fitting parameters \cite{bal2}. 
The parameter $B$ is related to the difference of the chemical
potentials of protons and neutrons in the fragment and 
depends on the temperature of the excited nucleus;
therefore it may reveal information about
the formation mechanism of the corresponding product.

Figure 1 shows the dependence of $Y_2/Y_1$ on $t_3$ for the entire mass 
region of product nuclei from proton-induced reactions
for different values of the difference in the neutron numbers
of considered pairs of targets  $\Delta N$. 
Similar dependences for deuteron-induced reactions are shown in Figure 2.
In both these figures, symbols show the measured data while lines
show their fit with formula (2). 

Tables 3 and 4 present the values of the fitting parameter $B$
for different combinations of targets pairs and for different
mass regions of product nuclei for proton- and deuteron-induced
reactions, respectively. 
Figure 3 shows the dependence of the parameter $B$ 
on the difference of neutron numbers in a pair of targets,
$\Delta N$, 
for different mass regions of products from proton-induced
reactions.

The value of the parameter $B$ increases linearly with increasing 
$\Delta N$.
$B$ also increases with increasing mass of the product nuclei.
The dependence of parameter $B$ 
on the difference of the neutron numbers in a pair of targets,
$\Delta N$, is fitted using the following formula:

\begin{equation}
B=k+d \Delta N,
\end{equation}
where $k=-0.036\pm0.01$ and $d=0.094\pm0.016$ 
for the mass region $7 \leq A \leq 30$,
$k=-0.0008\pm0.0001$ and $d=0.071\pm0.005$
for the mass region $40 \leq A \leq 80$, and
$k=-0.113\pm0.060$ and $d=0.033\pm0.008$ 
for the mass region $ A \geq 80$.
The value of the parameter $d$ changes with the mass number
of the products, and could be a factor in understanding 
the formation mechanism of the final nuclides.

From Tables 3 and 4, we see that for the production of 
$^{93-96}$Tc and $^{81-86}$Rb 
on the pair of targets $^{124}{\rm Sn}/^{112}{\rm Sn}$,
the parameter $B$ has values of
$1.07\pm0.32$ and $0.94\pm0.20$ for proton-induced reactions
and $1.10\pm0.40$ and $1.17\pm0.29$ for deuteron-induced reactions,
respectively. This agrees with similar values of $B$ of
$1.22\pm0.12$ and $1.23\pm0.13$
found in the literature for such products at a higher energy 
of 8.1 GeV \cite{bal2}. 
This allows us to conclude that residual products in this mass region
are produced via spallation processes of successive
particle evaporation.

\newpage
\begin{figure}[h]
\hspace*{-7mm}
\includegraphics[scale=0.35]{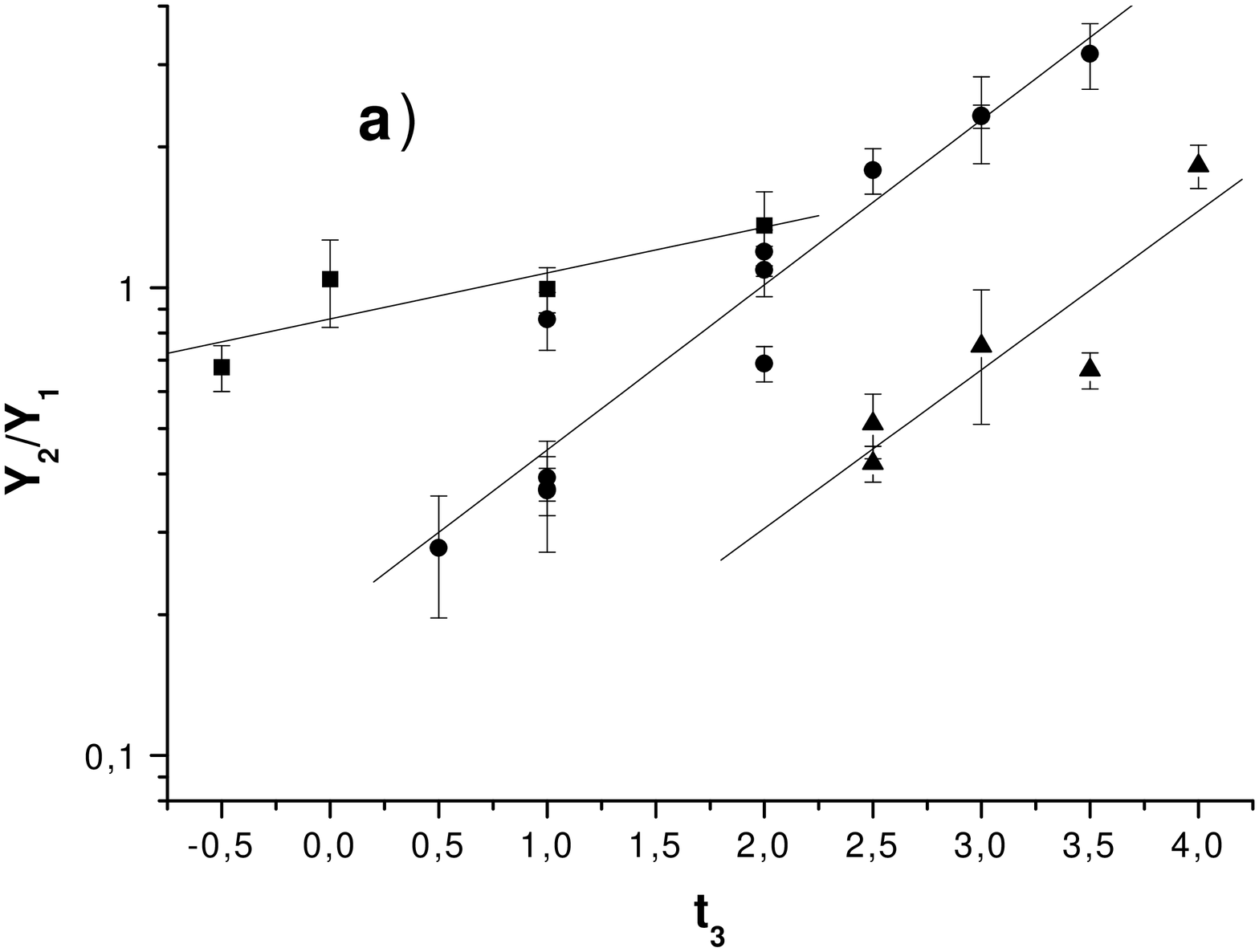}
\hspace*{-15mm}
\includegraphics[scale=0.35]{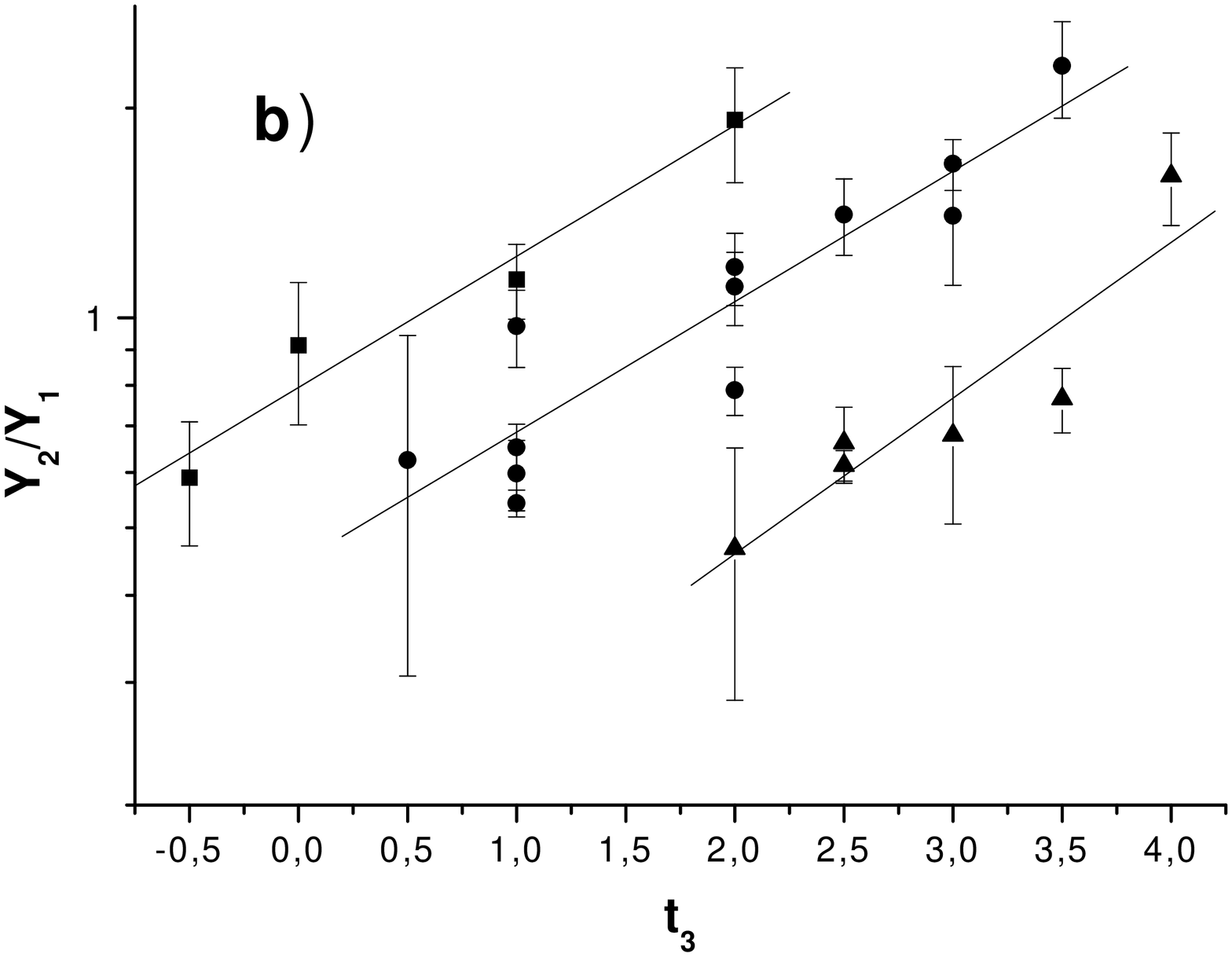}
\begin{center}
\hspace*{-5mm}
\includegraphics[scale=0.35]{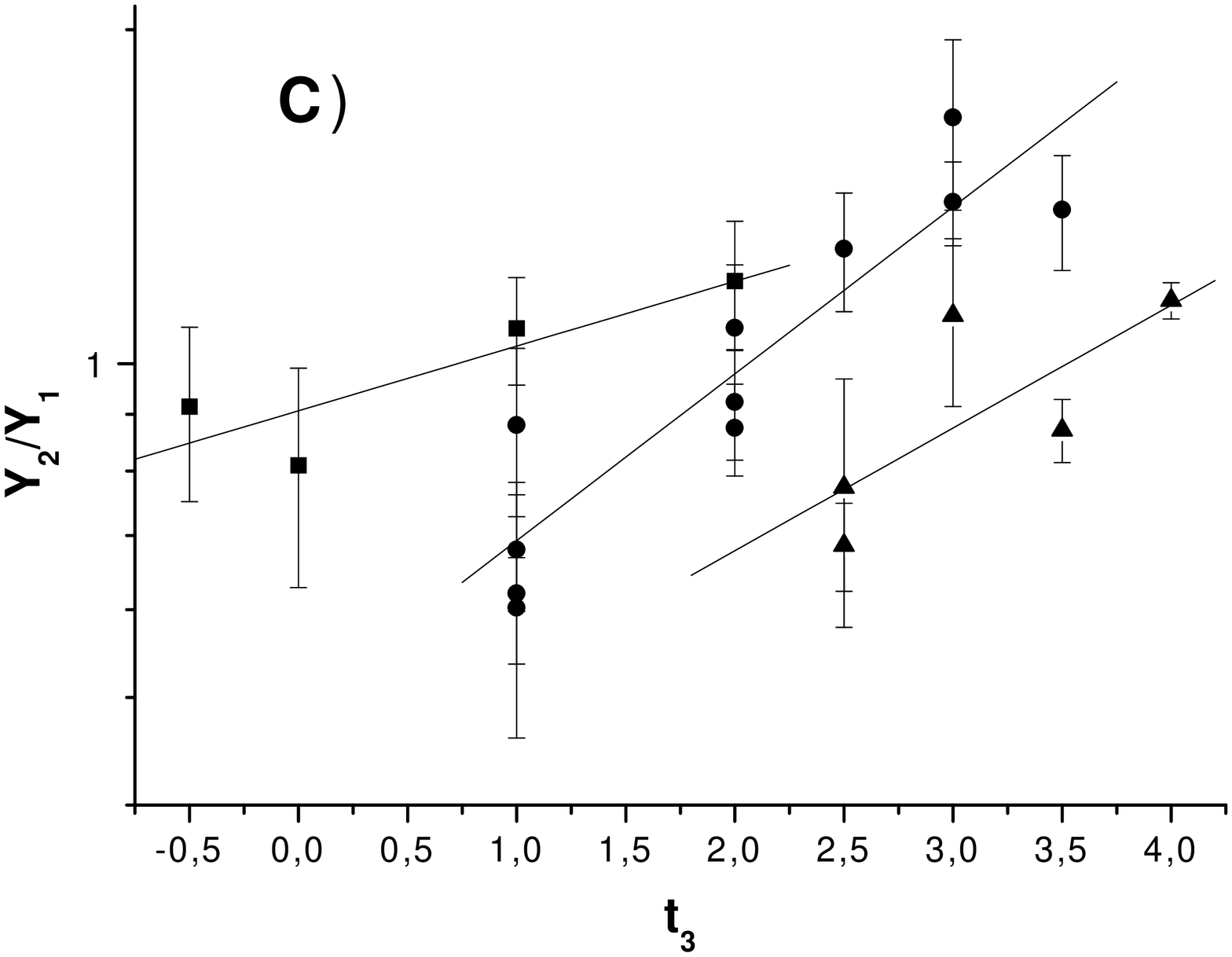}
\end{center}

\vspace*{-5mm}
\caption{
Ratio $R_{21}(t_3) = Y_2/Y_1$ versus
the isotopic-spin projection $t_3$ of products for 
different target pairs bombarded by protons: 
a) for $\Delta N$=12 (target pairs $^{124}$Sn/$^{112}$Sn);
b) for $\Delta N$=8 (target pairs $^{120}$Sn/$^{112}$Sn); and
c) for $\Delta N$=4 (target pairs $^{124}$Sn/$^{120}$Sn).
Symbols show measured yields of different products 
as following:
$\blacksquare$ --- for the mass region $7 \leq A \leq 30$; 
$\bullet$ --- for the mass region $40 \leq A \leq 60$;
and $\blacktriangle$ --- for the mass region $70 \leq A \leq 80$. 
Lines are results of fitting the data with formula (2).
}
\end{figure}

\begin{figure}[h]
\hspace*{-7mm}
\includegraphics[scale=0.35]{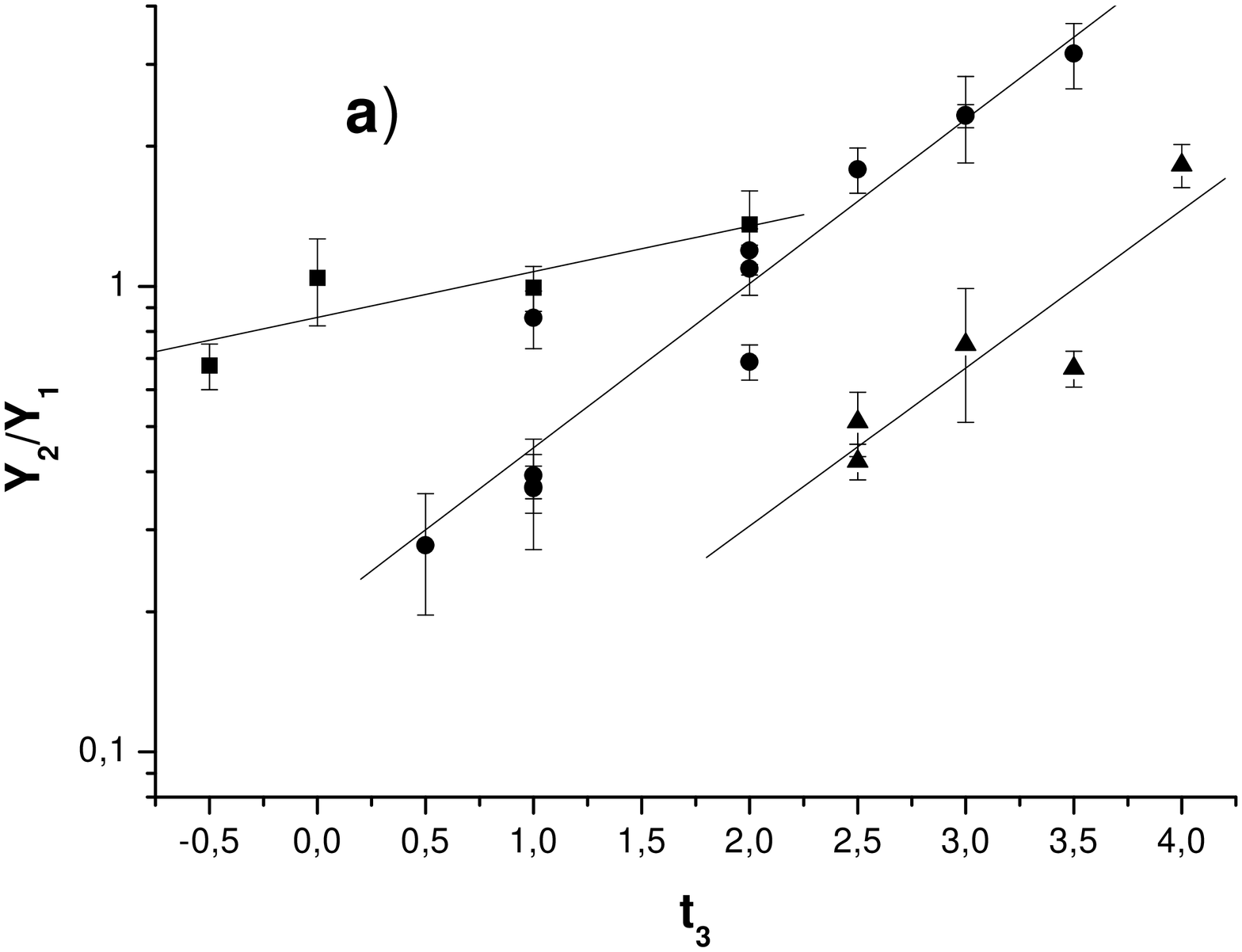}
\hspace*{-15mm}
\includegraphics[scale=0.35]{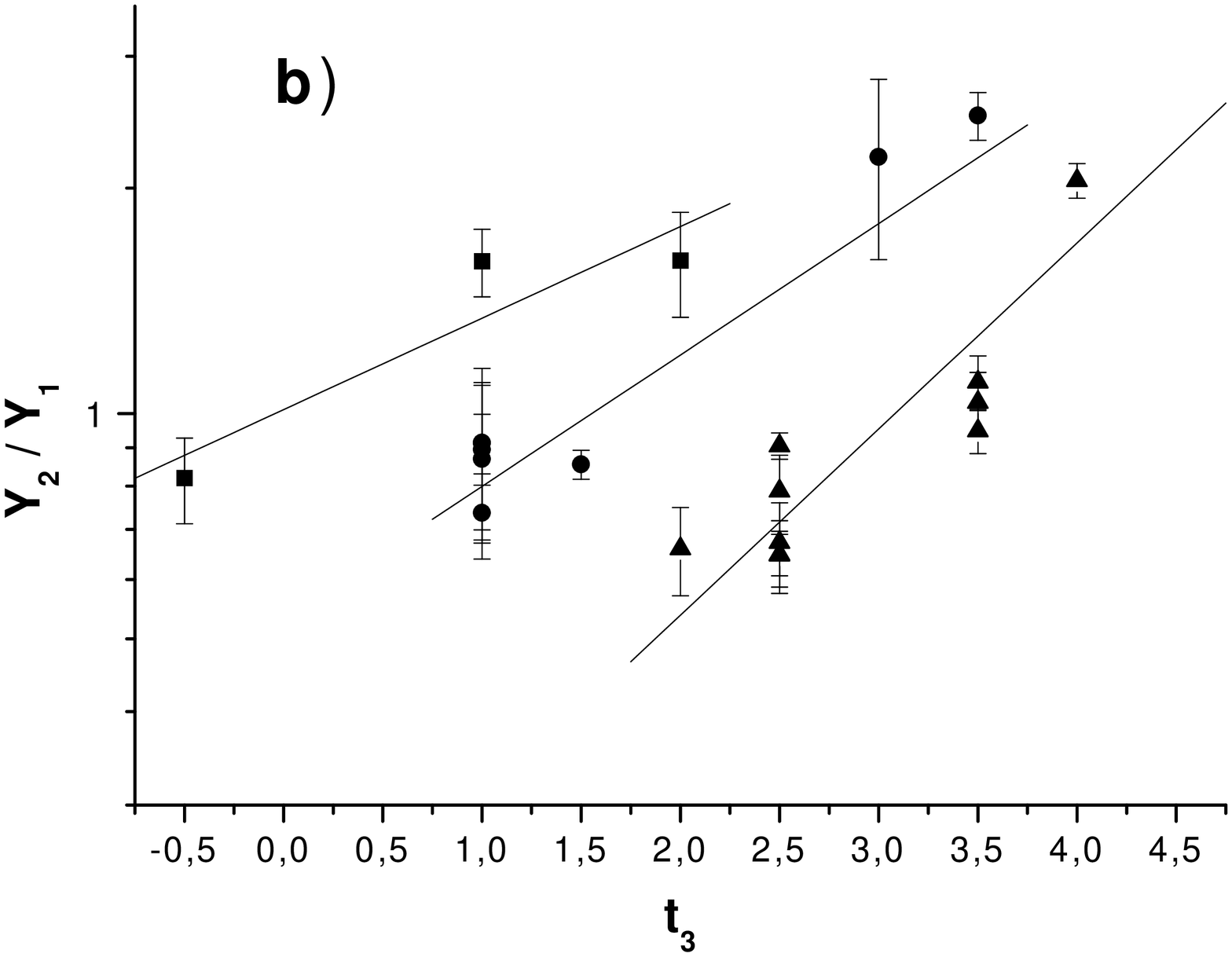}
\begin{center}
\hspace*{-5mm}
\includegraphics[scale=0.35]{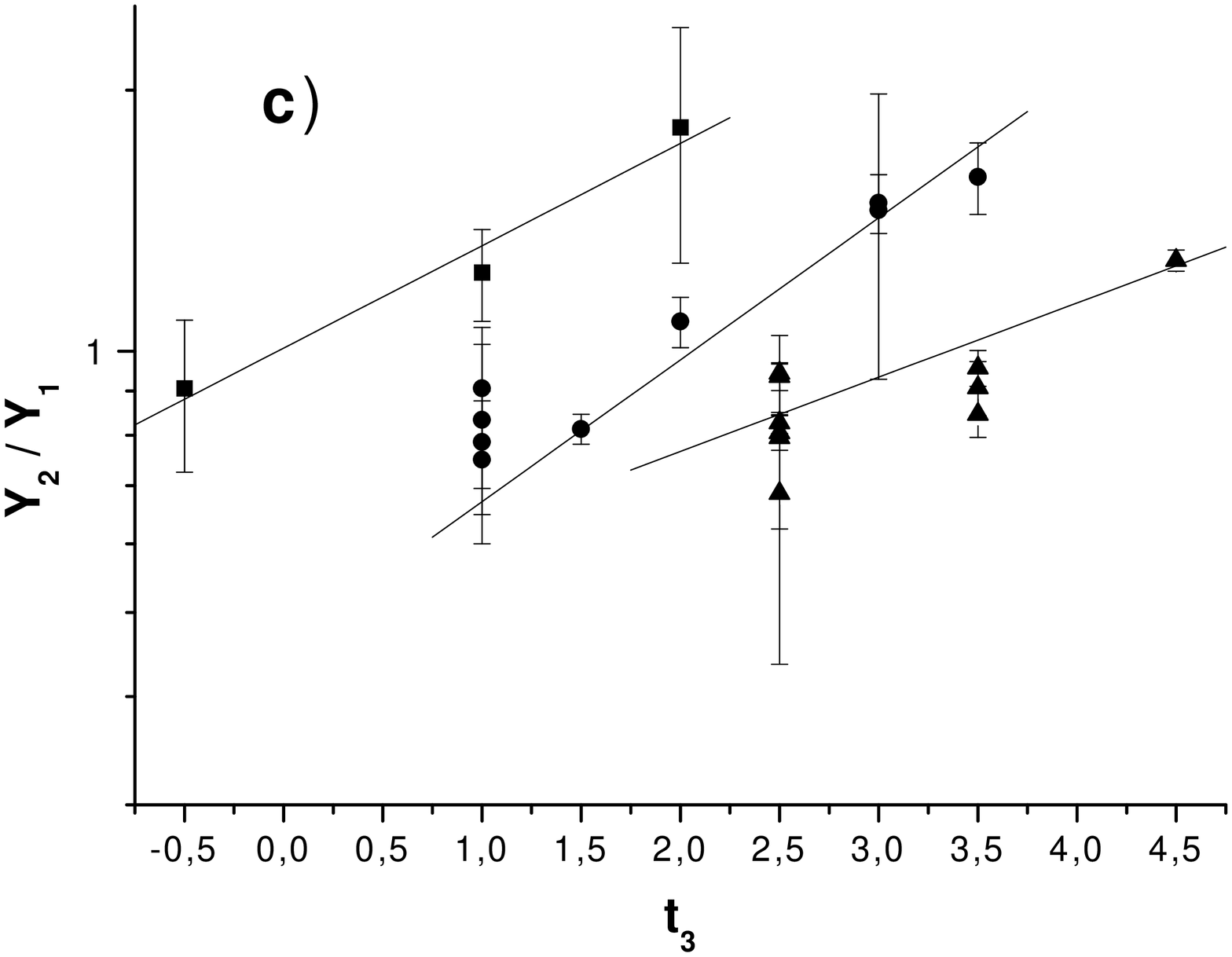}
\end{center}

\vspace*{-5mm}
\caption{
The same as in Fig.\ 1, but for deuteron-induced reactions.
}
\end{figure}
\clearpage

\begin{figure}[h]
\hspace*{-7mm}
\includegraphics[scale=0.35]{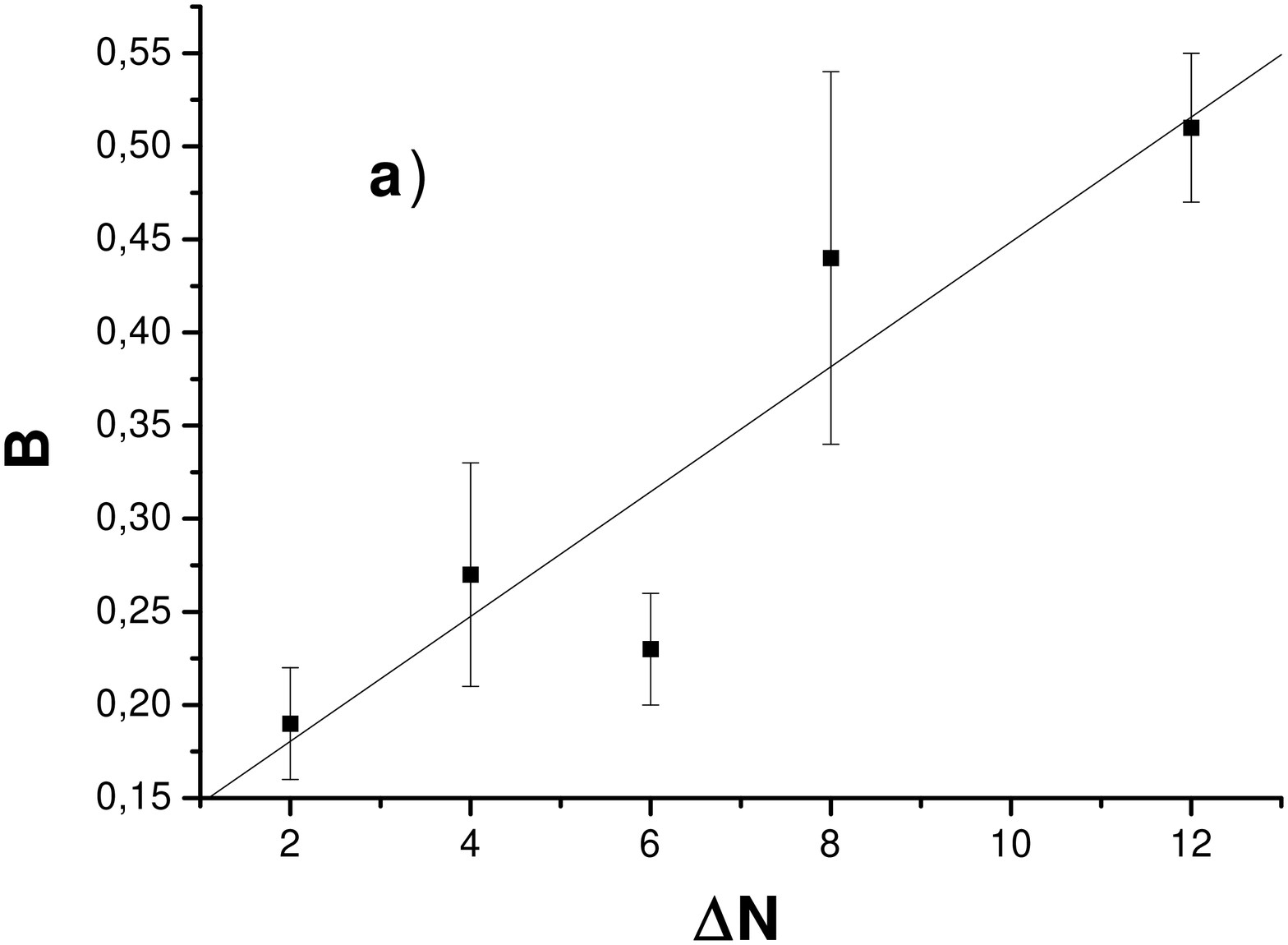}
\hspace*{-15mm}
\includegraphics[scale=0.35]{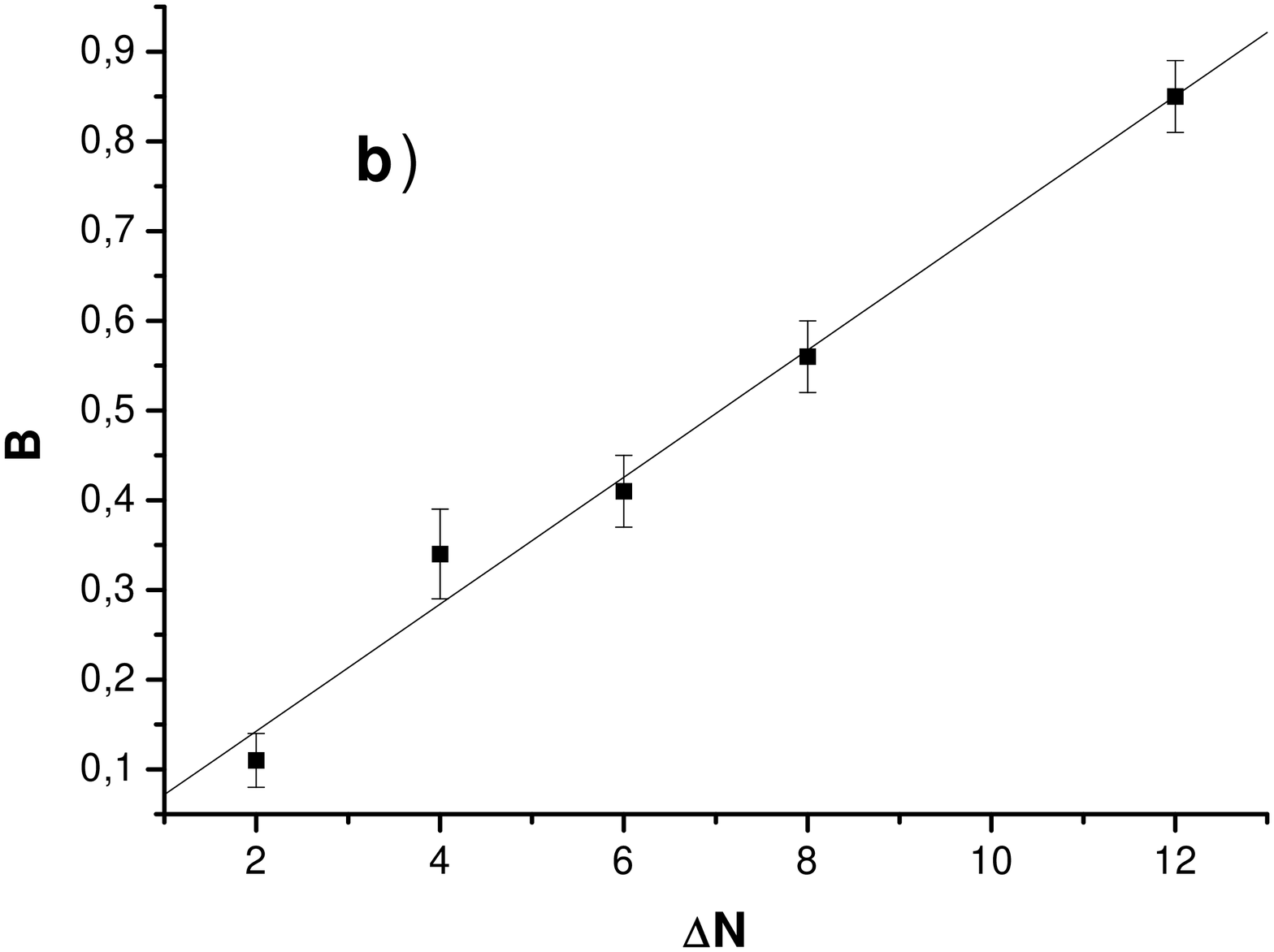}
\begin{center}
\hspace*{-5mm}
\includegraphics[scale=0.35]{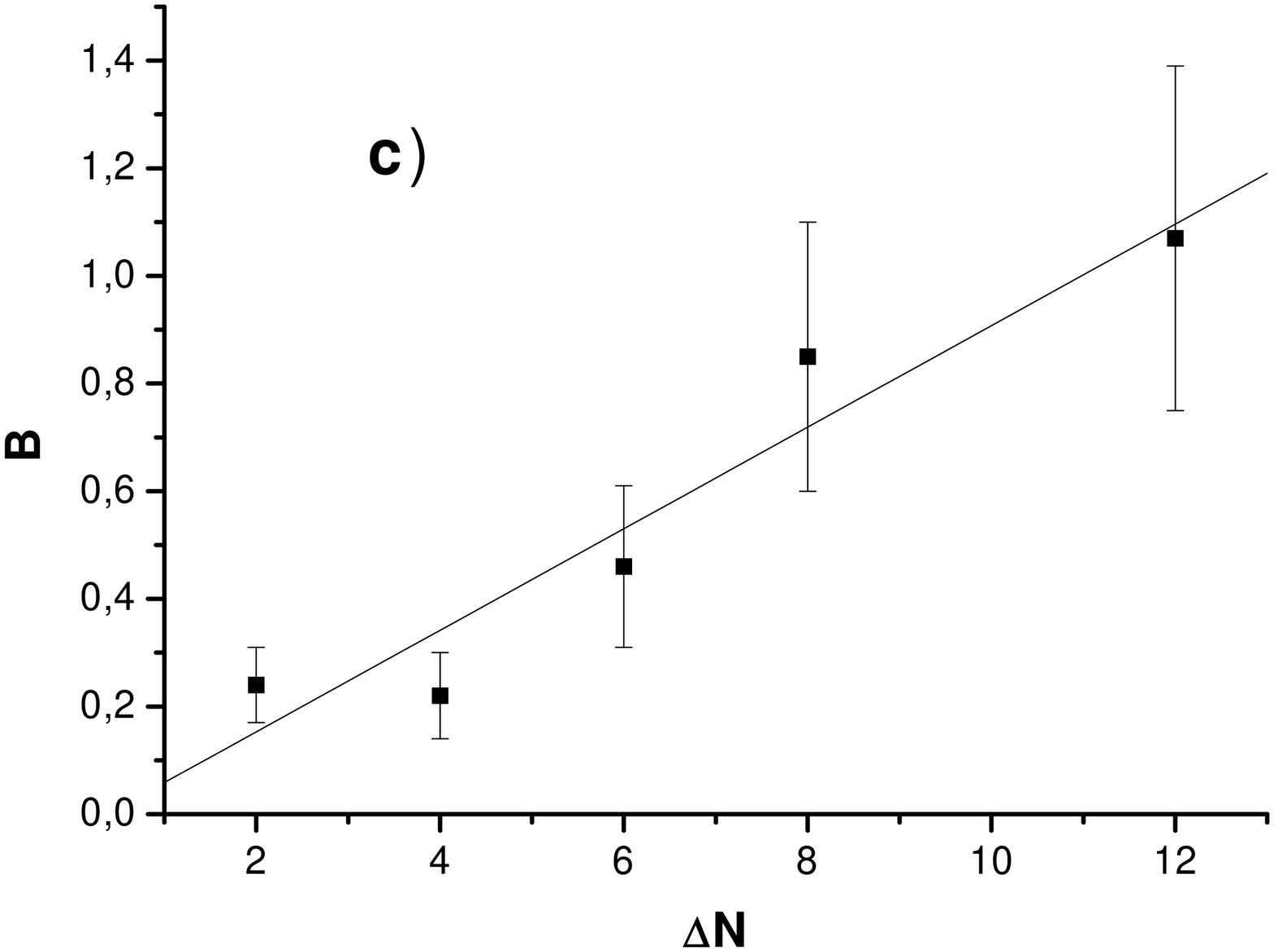}
\end{center}

\vspace*{-5mm}
\caption{
Parameter $B$ versus the difference in the excess of neutron
number ($\Delta N$) of targets for proton-induced reactions.
Symbols show values obtained by fitting the experimental
data with Eq. (2) as following:
a) for the product mass region $7 \leq A \leq 30$; 
b) for the mass region $40 \leq A \leq 80$; 
and c) for the mass region $ A \geq 80$.
Lines are results of fitting the parameter
$B$ with formula (3).
}
\end{figure}

\begin{center}
\textbf{Table 3}.
Mean values of the fitting parameter $B$ 
for different target pairs 
(with the difference in the excess neutron number of
$\Delta N$) bombarded by protons

\vspace*{5mm}
\begin{tabular}{|c|c|c|c|c|c|}  
\hline
Product nuclei&$\Delta N =2$&$\Delta N
=4$&$\Delta N =6$&$\Delta N =8$&$\Delta N =12$ \\ 
\hline
$7\leq A \leq 30$ &0.19$\pm$0.03& 0.27$\pm$0.06& 0.23$\pm$0.03&
0.44$\pm$0.10& 0.51$\pm$0.04 \\ 
\hline $40\leq A \leq 60$&
0.11$\pm$0.03& 0.34$\pm$0.05& 0.41$\pm$0.04& 0.56$\pm$0.04&
0.85$\pm$0.04 \\ 
\hline 
$70 \leq A \leq 80$ &0.18$\pm$0.05&
0.25$\pm$0.13& 0.36$\pm$0.09& 0.51$\pm$0.10& 0.78$\pm$0.21 \\
\hline 
$^{81-86}$Rb& 0.25$\pm$0.02& 0.32$\pm$0.04& 0.66$\pm$0.02&
0.62$\pm$0.15&0.94$\pm$0.20 \\ 
\hline
$^{93-96}$Tc& 0.24$\pm$0.07&
0.22$\pm$0.08& 0.46$\pm$0.15& 0.85$\pm$0.25& 1.07$\pm$0.32 \\
\hline
\end{tabular}
\vspace{1cm}

\textbf{Table 4}.
The same as in Table 3, but for deuteron-induced reactions

\vspace*{5mm}
\begin{tabular}{|c|c|c|c|c|}  
\hline
Product nuclei &$\Delta N =4$&$\Delta N
=6$&$\Delta N =8$&$\Delta N =12$ \\ 
\hline 
$7\leq A \leq 30$&
0.27$\pm$0.05& 0.26$\pm$0.05& 0.28$\pm$0.12& 0.55$\pm$0.07 \\
\hline 
$ 40 \leq A \leq 60$& 0.38$\pm$0.1& 0.41$\pm$0.09&
0.41$\pm$0.10& 0.78$\pm$0.14 \\ 
\hline 
$70 \leq A \leq 80$&
0.19$\pm$0.07& 0.5$\pm$0.1& 0.57$\pm$0.14& 0.77$\pm$0.19 \\ \hline
$^{81-86}$Rb& 0.30$\pm$0.08&       & 0.87$\pm$0.22&1.17$\pm$0.29\\ 
\hline
$^{93-96}T$c& 0.15$\pm$0.08& 0.48$\pm$0.30& 0.51$\pm$0.07& 1.10$\pm$0.40 \\
\hline
\end{tabular}
\end{center}

On the other hand, much smaller values of the fitting parameter $B$ 
in the mass region  $7\leq A \leq 30$
may point to a possible multifragmentation mechanism in the 
formation of these light fragments \cite{bal2,tsang1}.

A different situation may be seen in
the mass region $40\leq A\leq 60$, both for proton-
and deuteron-induced reactions.
The values of $B$ in this mass region 
is generally lower than 
for the heavy products $^{81-86}$Rb and $^{93-96}$Tc,
but higher than for light fragments with $7\leq A \leq 30$.
This may be understood if we assume that 
intermediate-mass nuclei are produced not only via
evaporation of particles (the spallation mechanism)
but also include a contribution from 
multifragmentation processes. 
This assumption is in agreement with results of our
earlier studies \cite{bal3} at bombarding proton energies of 0.66,
1.0, and 8.1 GeV:
We found that an observed
increase in the measured yields of intermediate-mass products 
can be described in the frameworks of 
the Intra-Nuclear Cascade (INC) model merged with SMM \cite{ban},
{\it i.e.}, by the INC+SMM model, which 
considers a contribution of multifragmentation to the formation of 
such intermediate-mass nuclei.

In the present work, we compare the measured cross sections
with predictions by the
FLUKA \cite{fluk1}, LAHET \cite{lahet},
CEM03 \cite{cem03}, and LAQGSM03 \cite{cem03} codes
(none of them considers the multifragmentation
mechanism of fragment production). The first three codes are
only applied to the proton-induced reactions, while LAGQSM03
is used for both protons and deuterons.
In order to compare the measured cumulative cross sections
with calculations, the corresponding
theoretical cumulative yields were estimated from 
the calculated independent cross sections. 

Figures 4, 5, and 6 show dependencies of ratios of theoretical to
experimental cross sections 
as functions on the product mass numbers for deuteron- and
proton-induced reaction, respectively. 
We see that, as a rule,
all models describe most of the measured cross sections 
of heavy and medium products within a factor of two.
Except for the CEM03 code, the agreement with the measured 
yields of light fragments
is much worse, where the other codes underestimate some
measured cross sections by up to two orders of magnitude 
and more. This could be related to the fact that all
the models used here do not consider multifragmentation.
But it is also true that they do not include simpler 
fission/fragmentation production mechanisms, either.

To have a better overall quantitative comparison of experimental data
with calculations, we have analyzed our data
using the mean deviation factor method suggested first
by R. Michel \cite{16}:

\begin{equation}\label{eqf}
\left<F\right>=10^{\sqrt{\left<(\log [ \sigma^{cal} / \sigma^{exp} ] )^2
\right>}}~,
\end{equation}
with its standard deviation
\begin{equation}\label{eqsf}
S\left(\left<F\right>\right)=10^{\sqrt{\left<\left(
\left|\log\left(
\sigma^{cal} / \sigma^{exp} \right)
\right|-\log(\left<F\right>)\right)^2\right>}}~,
\end{equation}
where $<>$ stands for averaging over all the products
included in the comparison. 

Values of the average deviation factor
$\left<F\right>$ and its standard deviation
$S\left(\left<F\right>\right)$ are listed in Table 5
for deuteron-induced reactions and in Table 6
for reactions with protons, respectively.

\begin{center}
\textbf{Table 5}. 
Mean deviations of product yields calculated by LAQGSM03
from the measured data
(parameters $<F>\pm S(<F>)$) for 
deuteron-induced reactions averaged over all
compared cross sections

\vspace*{5mm}
\begin{tabular}{|c|c|c|c|c|}  
\hline
&$^{112}$Sn&$^{118}$Sn&$^{120}$Sn&$^{124}$Sn\\ 
\hline
$<F>\pm S(<F>)$ &3.305$\pm$3.08 &2.04$\pm$1.69 & 2.96$\pm$2.75&2.41$\pm$2.75\\ 
\hline
\end{tabular}
\end{center}

\begin{center}
\textbf{Table 6}. 
Mean deviations of theoretical product yields
from the measured data (parameters $<F>\pm S(<F>)$) for 
proton-induced reactions averaged over all
compared cross sections

\vspace*{5mm}
\begin{tabular}{|c|c|c|c|c|}  \hline
 Models used&$^{112}$Sn&$^{118}$Sn&$^{120}$Sn&$^{124}$Sn
\\ \hline
LAHET &4.07$\pm$2.77 &3.49$\pm$2.40 &3.37$\pm$2.31 &3.61$\pm$2.56 \\ \hline
FLUKA &5.92$\pm$4.18 &7.84$\pm$5.19 &8.87$\pm$5.42 &6.97$\pm$4.29\\ \hline
LAQGSM03&5.10$\pm$3.86 &3.44$\pm$2.62 &3.09$\pm$2.22 &3.16$\pm$2.14 \\ \hline
CEM03 &3.66$\pm$3.02 &3.26$\pm$2.79 &4.04$\pm$3.29 &3.60$\pm$3.01 \\ \hline

\end{tabular}\vspace{1cm}
\end{center}

\vspace*{25mm}
\begin{figure}[h]
\begin{center}
\hspace*{-70mm}
\includegraphics[scale=0.68]{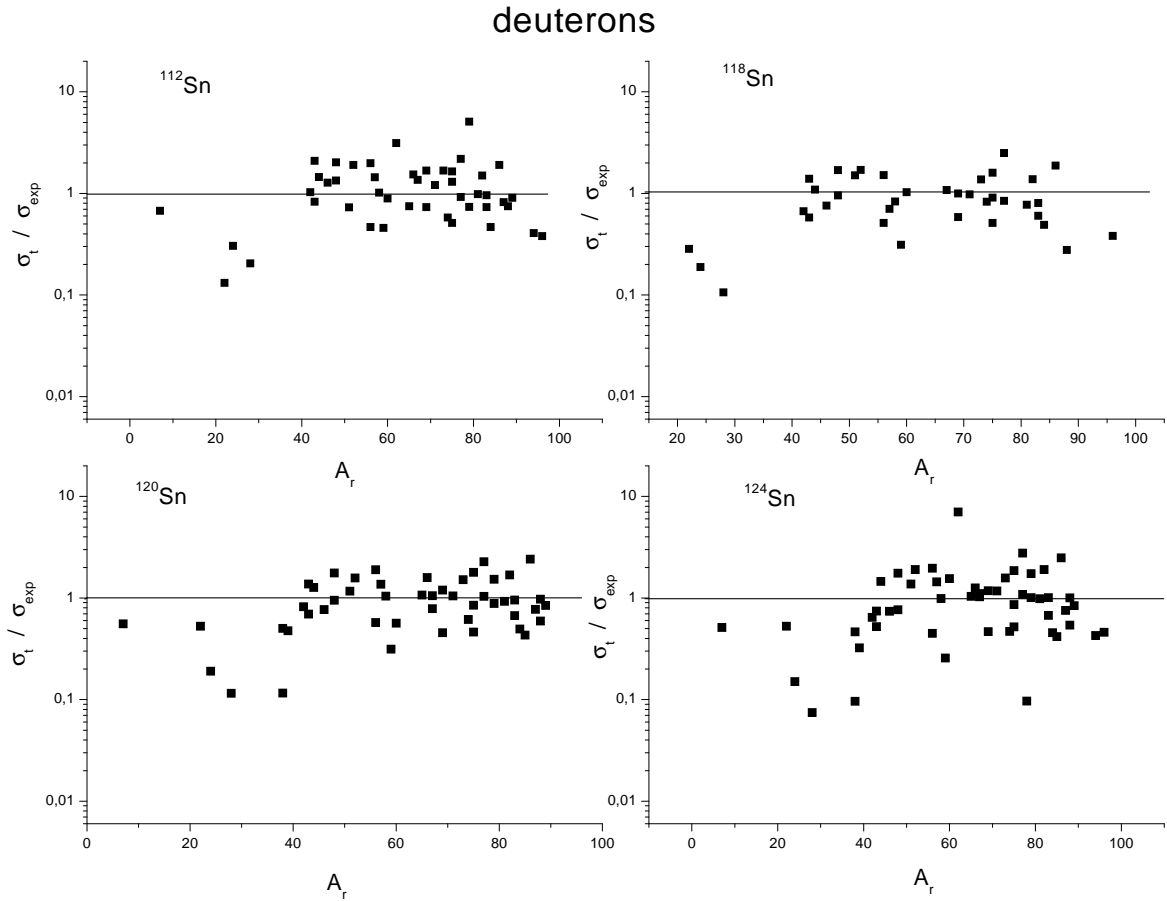}
\end{center}

\vspace*{-10mm}
\caption{
Dependence of the ratio of predicted
by LAQGSM03 and experimental
cross-sections on the mass number of products for deuteron-induced
reactions.
}
\end{figure}

\vspace*{-10mm}
\begin{figure}[h]
\begin{center}

\vspace*{-5mm}
\hspace*{-10mm}
\includegraphics[scale=0.85]{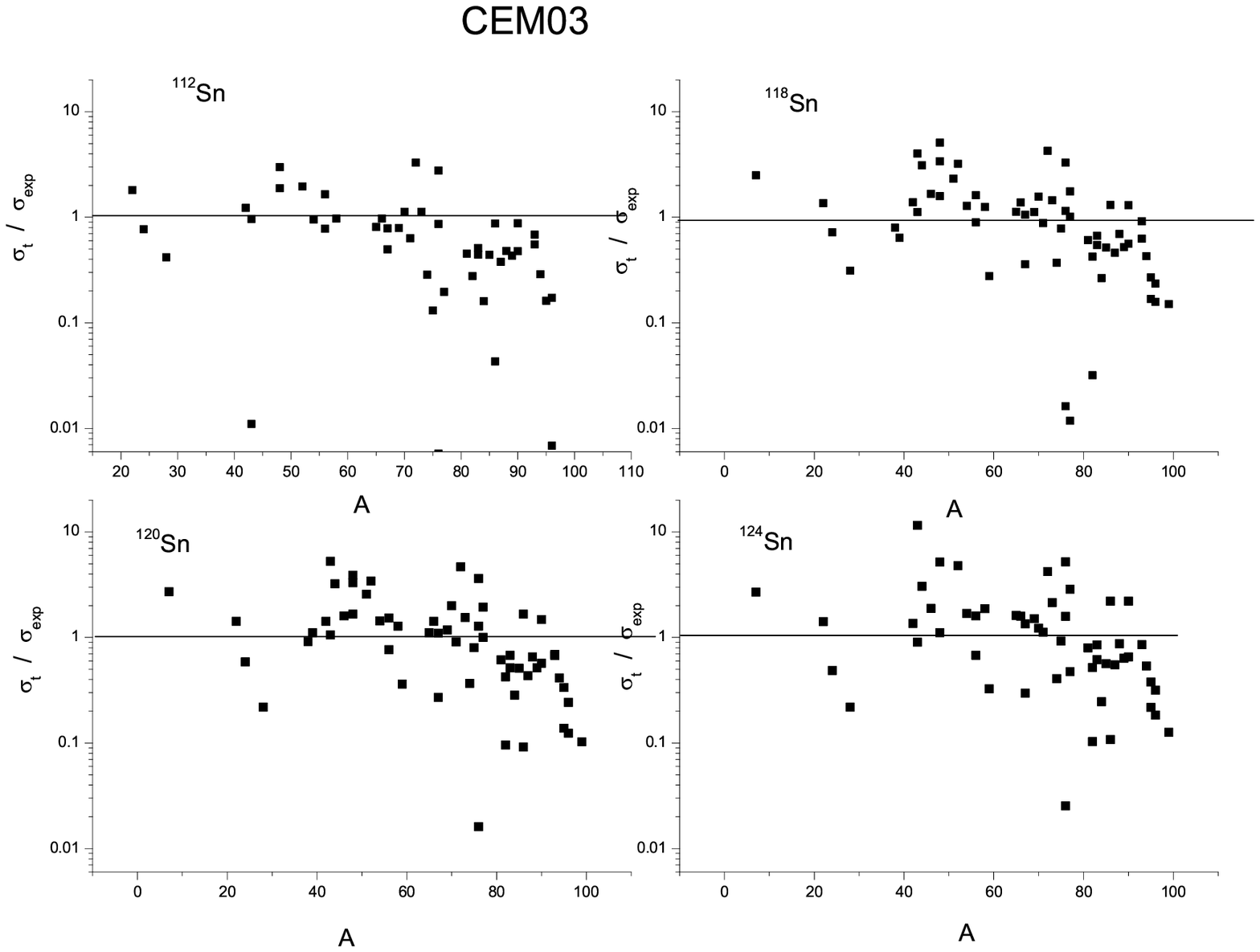}
\hspace*{-10mm}
\includegraphics[scale=0.85]{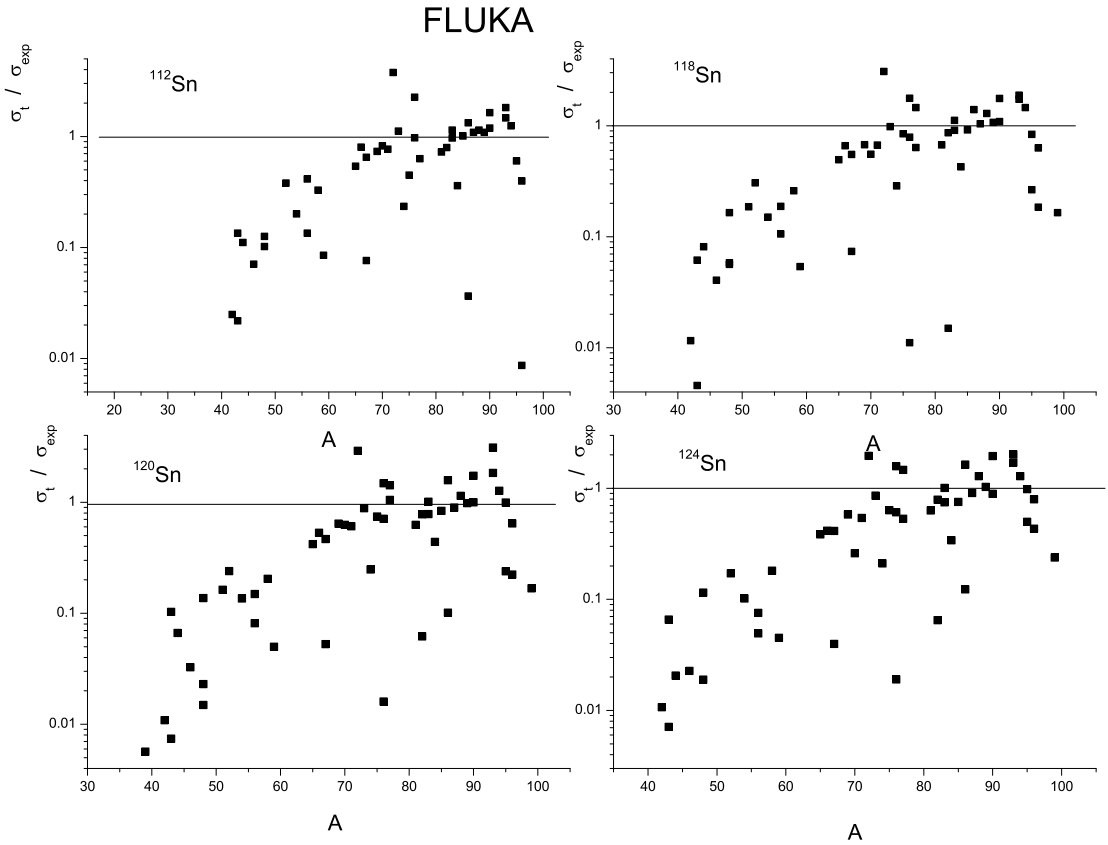}
\end{center}

\vspace*{-7mm}
\caption{
Dependences of ratios of predictions by  CEM03 and FLUKA
and experimental
cross-sections on the mass number of products for proton-induced
reactions.
}
\end{figure}

\vspace*{-10mm}
\begin{figure}[h]
\begin{center}

\vspace*{-10mm}
\hspace*{-10mm}
\includegraphics[scale=0.85]{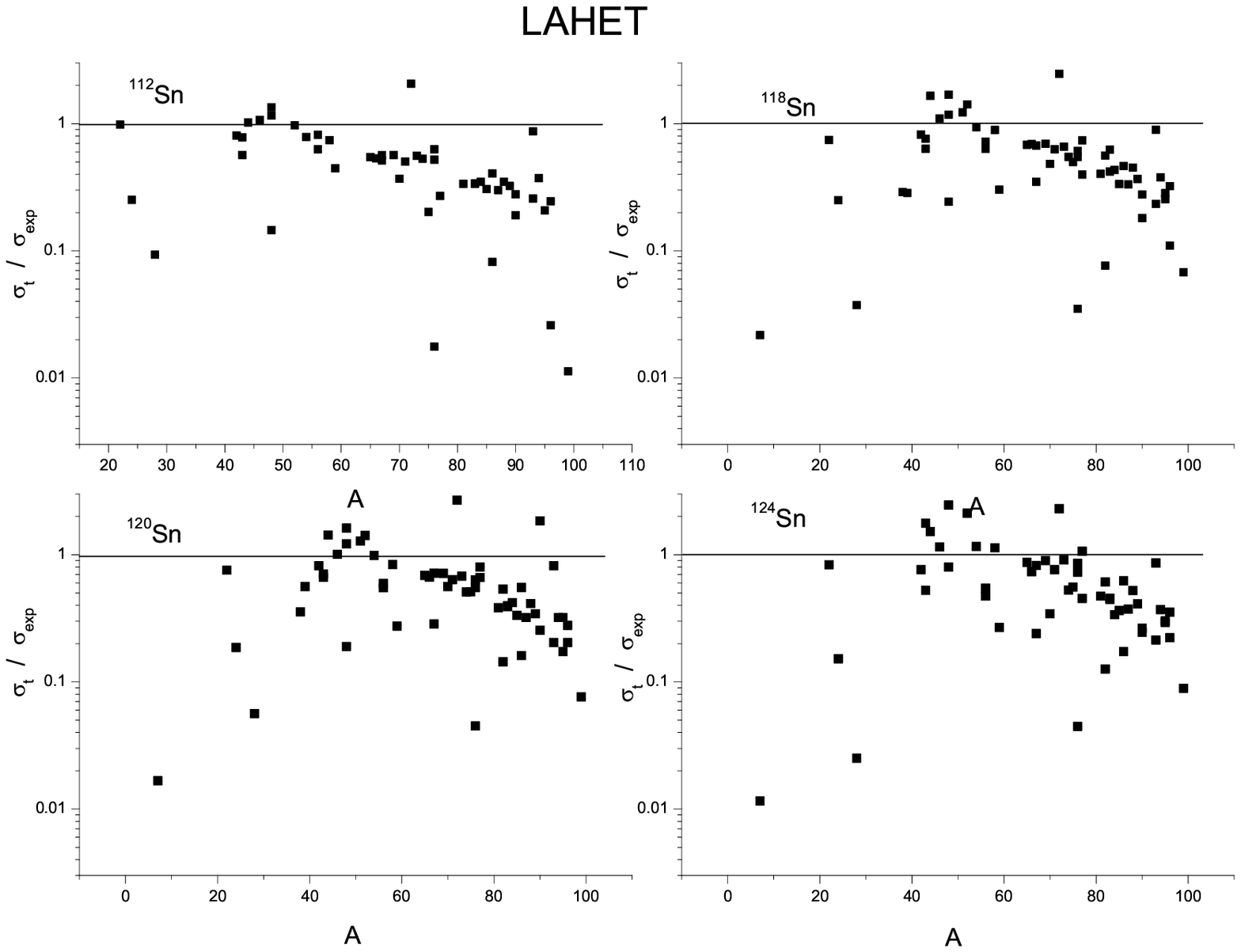}

\vspace*{38mm}

\hspace*{-74.7mm}
\includegraphics[scale=0.637]{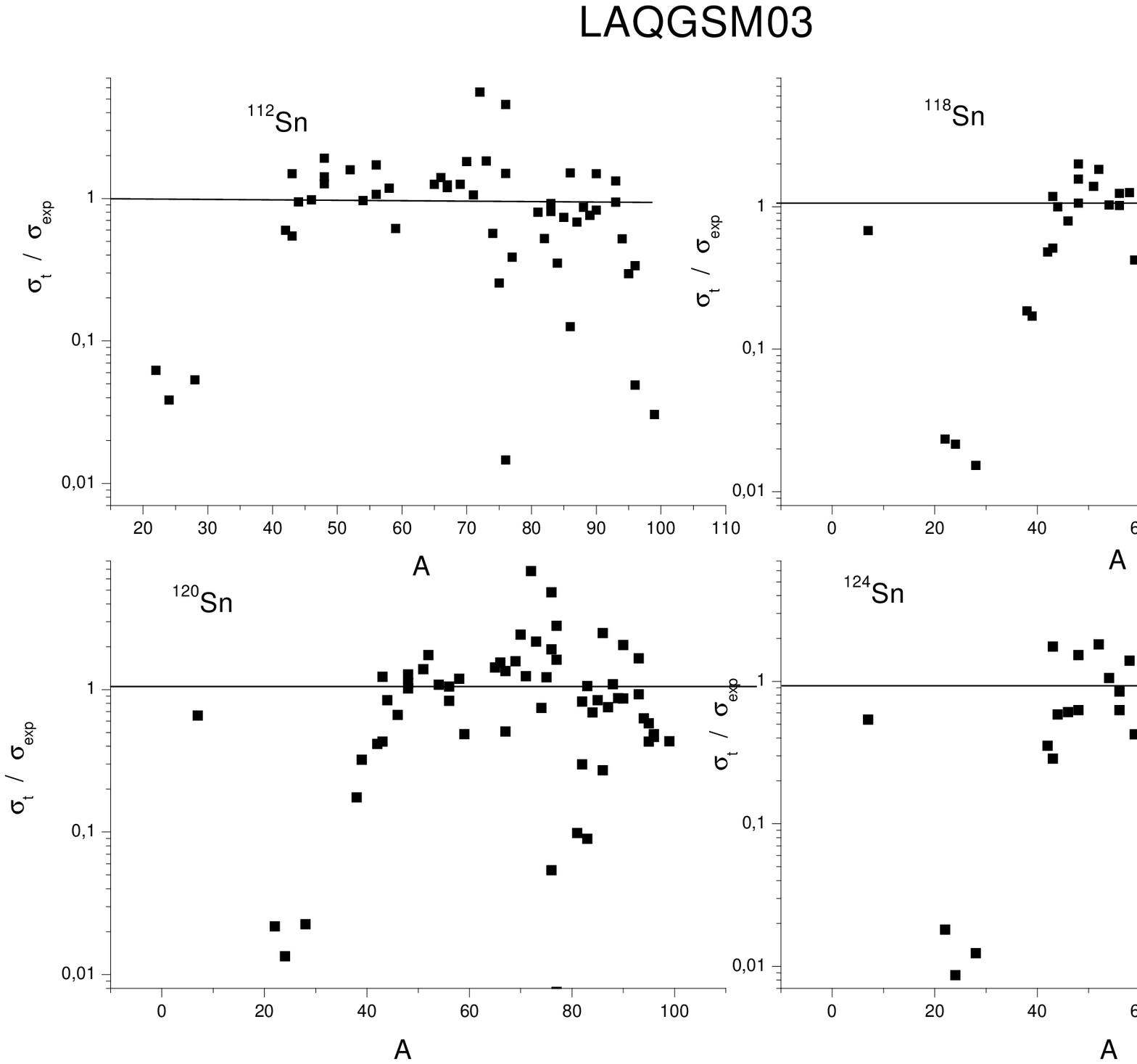}
\end{center}

\vspace*{-13mm}
\caption{
The same as in Fig.\ 5, but for LAHET and LAQGSM03.
}
\end{figure}
\clearpage

The present analysis points to a possible formation of light 
nuclides via multifragmentation, which would suggest a
``liquid-gas'' phase transition
taking place in hot nuclear matter formed by
irradiation of target nuclei with high-energy particles.
The intermediate-mass products are probably formed
mainly via evaporation, but some
contribution from multifragmentation
is also possible, according to our study.\\

{\bf Acknowledgments} \\

We thank Dr.\ A. J. Sierk for useful discussions and help.
The work was partially supported by the Advanced Simulation
Computing (ASC) Program at the Los Alamos National Laboratory
operated by the University of California for the U.~S.~Department
of Energy and by 
the Moldovan-US Bilateral Grants Program, CRDF Project MP2-3045-CH-02
and the NASA ATP01 Grant NRA-01-01-ATP-066.


\end{document}